\documentstyle[aps,preprint]{revtex}
\tightenlines

\newcommand{\beq}{\begin{equation}}
\newcommand{\eeq}{\end{equation}}
\newcommand{\bea}{\begin{eqnarray}}
\newcommand{\beas}{\begin{eqnarray*}}
\newcommand{\eea}{\end{eqnarray}}
\newcommand{\eeas}{\end{eqnarray*}}
\newcommand{\ba}{\begin{array}}
\newcommand{\ea}{\end{array}}
\def\ls{\mathrel{\lower4pt\vbox{\lineskip=0pt\baselineskip=0pt
           \hbox{$<$}\hbox{$\sim$}}}}
\def\gs{\mathrel{\lower4pt\vbox{\lineskip=0pt\baselineskip=0pt
           \hbox{$>$}\hbox{$\sim$}}}}

\begin{document}

\draft


\title{Possible astrophysical signatures of heavy stable neutral
relics in supergravity models}

\author{Rouzbeh Allahverdi~$^{1}$, Kari Enqvist~$^{2}$, and
Anupam Mazumdar~$^{3}$}

\address{$^{1}$ Physik Department, TU Muenchen, James Frank
Strasse, D-85748, Garching, Germany. \\
$^{2}$ Department of Physics and Helsinki Institute of Physics,
P. O. Box 9, FIN-00014, University of Helsinki, Finland.\\
$^{3}$ The Abdus Salam International Centre for Theoretical Physics, I-34100,
Trieste, Italy.}


\maketitle

\begin{abstract}
We consider heavy stable neutral particles
in the context of supergravity and show that a gravitationally suppressed
inflaton decay can produce such particles  in
cosmologically interesting abundances within a wide mass
range $10^3~{\rm GeV} \leq m_X \leq 10^{11}~{\rm GeV}$. In gravity-mediated
supersymmetry breaking models, a heavy particle can decay into its
superpartner and a photon-photino pair or a gravitino. Such decays
only change the identity of a possible dark matter candidate. However, for
$10^3~{\rm GeV} \leq m_X \leq 10^7~{\rm GeV}$, astrophysical bounds
from gamma-ray background and photodissociation of light elements
can be more stringent than the overclosure bound, thus ruling out the
particle as a dark matter candidate.
\end{abstract}

\vskip60pt

\section{Introduction}

Solving gauge hierarchy problem via hidden sector supersymmetry breaking
at a high scale due to non-perturbative dynamics which is then mediated to
the visible sector through gravity leads to a phenomenologically
successful prediction where sfermions and gauginos get a mass of order
electroweak scale \cite{nath}. In addition, the superpartner of a graviton,
the gravitino, also gets a mass of order $1$ TeV from super-Higgs
mechanism \cite{deser}. Although the gravitino interactions with matter are
suppressed by the Planck scale,  they can be generated in
a thermal bath from scattering of gauge and gaugino quanta with an
abundance given by \cite{ellis}\footnote{Recently, non-thermal
production of helicity $\pm 3/2$ gravitino \cite{maroto}, and
helicity $\pm 1/2$ gravitino \cite{kallosh} from time varying inflaton
oscillations have been considered. Helicity $\pm 1/2$ gravitino
for a single chiral multiplet is the superpartner of the inflaton known
as inflatino. The decay channels of inflatino have been discussed in
Ref.~\cite{rouzbeh}. Also, it has been suggested \cite{rouzbeh}, and
explicitly shown \cite{nps} that in realistic models with several chiral
multiplets helicity $\pm 1/2$ gravitino production is not a problem,
so long as the inflationary scale is sufficiently higher than the
scale of supersymmtery breaking in the hidden sector and the two
sectors are gravitationally coupled.}

\begin{equation}
\frac{n_{3/2}}{s} \approx 10^{-11}\frac{T_{\rm R}}{10^{11}~}\,,
\end{equation}
where $s$ defines the entropy density and $T_R$ denotes the reheating
temperature of the Universe in units of GeV. The gravitinos can then
be a source of potential trouble if they decay very late. The decay
rate of gravitino into gaugino and gauge, or, sfermion and fermion
quanta goes as $\Gamma \sim m_{3/2}^3/M_{\rm p}^2$, provided the
decay products have negligible mass compared to gravitino. This
posses a problem for nucleosynthesis for $m_{3/2}\sim {\cal O}({\rm TeV})$.
The gravitinos decay after nucleosynthesis and the decay products can change
the abundance of $^{4}\rm He$ and $\rm D$ by photodissociation. For
$100~{\rm GeV} \leq m_{3/2} \leq 1~{\rm TeV}$ a successful
nucleosynthesis limits the gravitino abundance to be
$n_{3/2}/s \leq (10^{-14}-10^{-12})$, which translates to
$T_{\rm R} \leq (10^7-10^9)~{\rm GeV}$ \cite{subir}.
This is the simplest illustration of late decaying particles in
cosmology which may release huge entropy while decaying. Various
cosmological and astrophysical observations put some useful constraints
on the abundance of late decaying particles. There are
many other examples within supergravity inflationary model which
has similar features as gravitinos, such as the moduli fields
and the dilaton \cite{beatriz}.

\vskip7pt

Massive particles which decay after nucleosynthesis may also distort
the spectrum of cosmic microwave background, or the gamma-ray
background. The exact astrophysical signature of such unstable relics
depends on their lifetime, thus some part of mass-lifetime parameter
space can be ruled out from the present experimental/observational
bounds \cite{kt}.

\vskip7pt

On the other hand, stable Weakly Interacting Massive Particles
(WIMPs), denoted here as $X$, generally respect the astrophysical
constraints, and can even account for the dark matter in the Universe
if they are produced in an interesting abundance. A famous example is
the Lightest Supersymmetric Particle (LSP) in supersymmetric
extensions of the standard model with unbroken $R$-parity.
If LSPs are created in a thermal bath,  LSPs pair annihilate to
lighter particles when the temperature of the Universe drops below
their mass. However, once the annihilation rate drops below the
Hubble expansion rate, the LSP comoving number density
freezes at its final value. The lower bound on the mass of such
species, in order not to overclose the Universe is a few GeV,
the so called Lee-Weinberg bound \cite{kt}, while unitarity provides a firm
upper bound on their mass $\leq 100$ TeV \cite{gk}.

\vskip7pt

The abovementioned bound on the mass of LSP can be evaded if there is no
initial thermal equilibrium. In recent years,
several mechanisms have been put forward for creating very heavy WIMPs,
even superheavy ones with a mass $m_X > 10^{10}$ GeV, in
cosmologically interesting abundances \cite{ckr1,ckr2,ckr3,cckr}.
For instance, the production of WIMPs could take place
during the phase transition from inflationary to the
radiation-dominated or matter-dominated phase \cite{ckr1}. The
quantum fluctuations of the field $X$ in a time-varying classical
gravitational background may lead to a significant production provided
$X$ is stable \cite{ckr1}. This mechanism is largely independent of
the nature of $X$ (boson or fermion) and its coupling to other
fields. It also works in a variety of inflationary models \cite{cckr}
but abundances close to the dark matter abundance are created for
$10^{11}~{\rm GeV} \leq m_X \leq 10^{13}~{\rm GeV}$ and $T_R \simeq
10^9$ GeV.

Another possibility is to create a bosonic field $X$ during the rapid
oscillations of the inflaton field from vacuum fluctuations via a
coupling $g^2 \phi^2 X^2$, where $g \sim {\cal O}(1)$
\cite{ckr2}. However, one then requires $m_X \geq 10 m_\phi$ and a low
reheat temperature, e.g. $T_{\rm R} \sim 10^2$ GeV for $m_\phi \sim 10^{13}$
GeV, which makes the mechanism model-dependent.

If the plasma of the inflaton decay products has an instantaneous
temperature $T \sim {(T^{2}_{\rm R} H M_{\rm Planck})}^{1/4}$ for
$H \geq \Gamma_{\rm d}$, where $\Gamma_{\rm d}$ is the inflaton decay
rate, it can be much higher than the reheat temperature of the
Universe \cite {kt}. Then particles of mass $m_X > T_{\rm R}$ and with
gauge strength interactions can be
produced from the annihilation of the relativistic particles in the
thermal bath and their abundance freezes at its final value once
temperature becomes sufficiently low \cite{ckr3}.

\vskip7pt

However, none of the above mentioned
scenarios for creating WIMPs in the mass range $\sim (10^{10}-10^{13})$ GeV
has been actually realized in a supersymmetric set-up. For instance,
the identity of $X$ is largely unknown apart from the fact that it should be a Standard Model (SM)
gauge singlet. However, if one assumes that
$X$ comes from a hidden sector which interacts only
gravitationally, they could act as a
candidate for dark matter. Within supersymmetry, a
heavy particle $X$ is accompanied by its superpartner $\tilde X$ with an
almost degenerate mass. This is due to the fact that supersymmetry
in the visible sector is broken at a scale $\sim 1~ {\rm TeV}$. Moreover,
in supergravity $X$ and $\tilde X$ have gravitationally suppressed
couplings to other fields, among which most notable are the inflaton
$\phi$, its superpartner inflatino $\tilde \phi$, and the gravitino. This
opens up another possibility for creating $X$ via gravitationally
suppressed decay of the inflaton. Also, couplings of $X$; $\tilde X$ to
$\phi$; $\tilde \phi$ and the gravitino, together with a small mass
difference between $X$ and $\tilde X$, can result in the decay of $X$
($\tilde X$) into its superpartner and a photon-photino pair, or a
gravitino. These decay channels however preserve any symmetry which
is necessary to forbid $X$ and $\tilde X$ directly decaying into the
SM fields.
The startling point is that while decaying into its own superpartner,
the process changes the identity of a possible dark matter
candidate. However, while doing so the process releases some energy
into a cosmic thermal bath. In this paper we discuss the astrophysical
bounds on such decays for gravitino masses ranging from $100$ GeV to
$1$ TeV.

\section{Production of $X$ from its coupling to the inflaton sector}

We may consider a complex scalar field $X$ and its fermionic partner
$\tilde{X}$ with a superpotential mass term
$W \supset (1/2) m_X {\bf X} {\bf X}$. In the limit of unbroken
supersymmetry $X$ and $\tilde X$ have a common mass $m_X$. However,
hidden sector supersymmetry breaking generally results in the soft
mass term;
$m^{2}_{3/2} |X|^2$, and the $A$-term; $A m_{3/2} m_X X X + {\rm h.c}$.
The former elevates the masses of both the components of $X$
above the mass of $\tilde X$, while the latter one results in the splitting
of $X$ into two mass eigenstates $X_1$ and $X_2$, where
we may take  $m_{X_1} > m_{X_2}$. When $m_X \gg m_{3/2}$, only the
contribution of $A$-term has any significance and yields

\begin{equation}
m_{\tilde X} \approx m_X\,;~~~~m_{X_1} + m_{X_2} \approx 2 m_X \,.
\end{equation}
We denote the inflaton multiplet by $\Phi$;  it comprises of the
inflaton $\phi$ and the inflatino $\tilde \phi$. Around the global
minimum of the potential, the inflation sector superpotential can be
approximated as $W \supset (1/2) m_\phi {(\Phi-v)}^2$, where $v$ is the
vev of $\phi$ at the minimum. The
new inflationary and hybrid inflationary models usually give rise to
such a non-zero
VEV for either inflaton or some auxiliary field. There may
exist a superpotential term $h_X \Phi {\bf X} {\bf X}$ which couples
$\Phi$ and ${\bf X}$ multiplets \footnote{Note that a superpotential term
${\bf X} \Phi \Phi$ which is linear in ${\bf X}$ provides a decay
channel for $X$ and $\tilde X$ thus ignored in our discussion.}. In
general they can also be coupled via higher dimensional Planck
suppressed terms. For example, consider the minimal supergravity where
the scalar potential is given by \cite{nath}

\beq
V = e^G \left ( {\partial G \over \partial {\varphi}_i} {\partial G
\over \partial {\varphi}^{*}_{i}} - {3 \over M^{2}_{\rm Planck}}
\right ).
\eeq
Here $G$ is the k\"ahler function defined by

\beq
G = {{\varphi_i} \varphi^{*}_{i} \over M^{2}_{\rm Planck}} + {\rm
ln} \left ({{|W|}^{2} \over M^{6}_{\rm Planck}} \right ),
\eeq
where $\varphi_i$ are scalar fields in the theory. There also exists a
term in the Lagrangian

\beq
e^{G/2} \left ({\partial G \over \partial {\varphi}_i} {\partial G \over
\partial {\varphi}_j} - {\partial ^2 G \over {\partial
\varphi_i} {\partial \varphi_j}} \right ) \bar{\tilde{\varphi}_i}
{\tilde{\varphi}_j} + {\rm h.c.}, 
\eeq
with $\tilde {\varphi}_{i}$ being the fermionic partner of
$\varphi_i$. For our choice of superpotential the following
terms can then be identified in the Lagrangian
\beq
\label{lagrangian}
{v m_X \over M^{2}_{\rm Planck}} \phi {\tilde X} {\tilde X}\,;~~~
{v m_X \over M^{2}_{\rm Planck}} {\tilde \phi} {\tilde X} X\,;~~~
{v m_\phi m_X \over M^{2}_{\rm Planck}} \phi X X \,.
\eeq
and the coupling $h_X$ recognized as

\begin{equation}
\label{coupling}
h_X = \frac{v m_X}{M^{2}_{\rm Planck}}\,,
\end{equation}

\subsection{Production by direct inflaton decay}

The terms in Eq.~(\ref{lagrangian}) result in $X$ ($\tilde X$) production
from inflaton decay if $m_X < m_\phi/2$, with a rate
$\Gamma_X \sim h^{2}_{X} m_\phi/{8 \pi}$. The total inflaton decay rate
is given by $\Gamma_d \sim \sqrt{8\pi/3}~T^{2}_{\rm R}/M_{\rm
Planck}$, while
the inflaton number density at the time of decay is given by
$n_\phi \sim T^{4}_{\rm R}/m_\phi$. This constrains the overall
coupling to
\beq
h^{2}_{X} \leq 32 \pi \sqrt{8 \pi \over 3} {T_{\rm R} \over M_{\rm Planck}}
{10^{-9} \over m_X}\,,
\eeq
where $m_X$ is in units of GeV. This is required due to the fact that
the produced $X$ ($\tilde X$) must not overclose the Universe which,
for $\Omega_X \leq 0.3$ and $H_0 = 70$ km ${\rm sec}^{-1}
{\rm Mpc}^{-1}$, reads

\beq
{n_X \over n_\gamma} \leq {4 \times 10^{-9} \over m_X},
\eeq
when $m_X$ is expressed in units of GeV. In the particular case when
$\Phi$ and ${\bf X}$ multiplets are gravitationally coupled one finds

\beq
m^{3}_{X} \left({v \over M_{\rm Planck}}\right)^2 \leq
10^{-7} M_{\rm Planck}~T_{\rm R}.
\eeq
If $v \simeq M_{\rm Planck}$, then $X$ ($\tilde X$) within the mass range
$10^3~{\rm GeV} \leq m_X \leq 10^7~{\rm GeV}$ can be produced in
interesting abundances for a range of reheat temperatures
$1~{\rm MeV} \leq T_{\rm R} \leq 10^{10}~{\rm GeV}$.

If the inflaton VEV is lowered down to
$v/M_{\rm Planck} \simeq 10^{-6}$, then $X$ ($\tilde X$)
with a mass range $10^7~{\rm GeV} \leq m_X \leq 10^{11}~{\rm GeV}$ can
become
a dark matter candidate for the same range of reheat temperature
\footnote{It is evident from the overclosure bound that $h_X$ needs to
be very small. This may look ad-hoc for a superpotential coupling between
$\Phi$ and ${\bf X}$, while a non-renormalizable gravitationally
suppressed coupling in the context of supergravity offers a natural
explanation for such a small coupling if $X$ belongs to a hidden sector.}.

\subsection{Production from the thermal bath}

Another possibility for production of $X$ is from its indirect
coupling to the
thermal bath via the inflaton sector. The simplest situation will be
for the case when inflaton is coupled to two fermions (hence
the inflatino is coupled to a fermion-boson pair). As an example
consider the $\phi {\tilde \gamma} {\tilde \gamma}$ and ${\tilde \phi}
{\tilde \gamma} \gamma$ couplings\footnote{The situation when $\gamma$
and $\tilde \gamma$ are replaced by a light particle and its
superpartner is similar.}. These couplings are supersymmetric
partners and the latter one is responsible for a Hubble-induced
gaugino mass term $\propto H {\tilde \gamma} {\tilde \gamma}$. Then
$X$ ($\tilde X$) can be produced in a thermal bath from the s-channel
diagram which includes $h \phi {\tilde \gamma} {\tilde \gamma}$ ($h
{\tilde \phi} {\tilde \gamma} \gamma$) and $h_X \phi {\tilde X}
{\tilde X}$ ($h_X {\tilde \phi} {\tilde X} X$) couplings at the first
and second vertices respectively. Assuming that $m_X \leq T_{\rm R}$,
the rate
for $X$ ($\tilde X$) production is given by

\beq
\Gamma_X \sim {h^2 h^{2}_{X} \over 16 \pi^2 m^{2}_{\phi}} T^{3}_{\rm R},
\eeq
where $m_\phi \approx m_{\tilde \phi}$, since hidden sector
supersymmtery breaking only results in $|m_\phi - m_{\tilde \phi}| \ll
m_\phi$. The coupling $h$ obeys the following relationship

\beq
h^2 {m_\phi \over 8 \pi} \sim \sqrt{8 \pi \over 3} {T^{2}_{\rm R}
\over M_{\rm Planck}},
\eeq
and the number density of created $X$ ($\tilde X$) particles follows

\beq
{n_X \over n_\gamma} \sim {h^{2}_{X} \over 2 \pi} {\left ({T_{\rm R} \over
m_\phi} \right )}^3.
\eeq
Here the main contribution to $X$ ($\tilde X$) production occurs
at $H \simeq \Gamma_{\rm d}$. In order for $X$
($\tilde X$) not to overclose the Universe it is necessary to have
$\Omega_X \leq 0.3$, which implies

\beq
h^{2}_{X} \leq {3 \times 10^{-8} \over m_X} {\left ({m_\phi \over T_{\rm
R}} \right)}^3.
\eeq
We always have $\Gamma_X < H$ since $h^{2}_{X} T^{3}_{\rm R} <
m^{3}_{\phi}$ (note that for perturbative inflaton decay $T_{\rm R}
< m_\phi$). This ensures that $X$ ($\tilde X$) will never reach
equilibrium with the thermal bath. However, the bound on $h_X$ from
(8) is much stronger than the bound in (14). This implies that $X$
($\tilde X$) production from thermal bath is always subleading with respect to
production by direct inflaton decay in order for the latter not to
saturate the overclosure bound.

\vskip7pt

It is interesting that direct inflaton decay in supergravity may produce
$X$ ($\tilde X$) in a wide range of mass with a cosmologically
interesting abundance. This is particularly significant as some of
the proposed mechanism for creating heavy WIMPs might
not work in a supersymmetric set-up.
For example, it is known that scalar fields, and also fermions
if allowed by symmetry considerations, acquire a
Hubble-induced mass term during and after inflation in supergravity
models. This suggests that production of such particles from a
time-varying gravitational backgrouned might not be possible at all
unless such Hubble-induced term is small, or cancelled for a
particular choice of superpotential.

\section{Very Late decay of $X~(\tilde X)$}

There exist strong astrophysical bounds on the late decay of
particles to photons or charged particles. For a detailed study of
various constraints on such decays we refer the reader to
Refs.~\cite{subir1,kr}. Here we briefly mention the relevant bounds.
For a late decay such as
$10^5 ~ {\rm s} \leq \tau \leq 10^{13} ~ {\rm s}$, where
$\tau$ denotes the lifetime of a particle, the decay products can
alter the chemical potential of the microwave background photons. Note
that recombination occurs at $10^{13}$ s. The reason is that
photon number changing interactions such as
$e ~\gamma \rightarrow e ~ \gamma ~ \gamma$ go out of equilibrium at
$t > 10^{5}$ s thus inducing an effective chemical potential for
the microwave background photons.
The chemical potential has been given in \cite{dolgov} for the
case that decay photons thermalize instantly and the current experimental
bound is \cite{mather}

\begin{equation}
\label{const1}
\mu =\frac{\rho_{\rm d}}{\rho_{\gamma}} \leq 3.3 \times 10^{-4}\,,
\end{equation}
where $\rho_{d}$ and $\rho_{\gamma}$ denote the energy density in
the decay and background photons respectively. If the decay occurs
after recombination and before the present era, then the non-thermal
decay photons may be directly visible today if
the optical depth back to the decay epoch is small enough. Then the
main astrophysical constraint is that the flux of decay photons
must not exceed that of the observed differential photon
flux which is given by \cite{kt} \footnote{Decay photons generally
trigger an
electromagnetic shower, since their scattering off the microwave
background photons can create $e^+e^-$ pairs, which themselves undergo
Compton scattering off photons. Moreover, $\gamma-\gamma$ scattering
redistribute photon energy. Therefore one must take these effects into
account
in order to compare with the observed photon flux, as done in
\cite{subir1,kr}.}

\begin{equation}
\label{const}
{\cal F}_{\gamma}(\rm E)
\leq \frac{\rm MeV}{\rm E} ~ {\rm cm^{-2} ~ sr^{-1} ~ s^{-1}}\,,
\end{equation}
where $\rm E$ is the photon energy and the observed photon flux is
denoted by ${\cal F}_\gamma$.

There are also bounds on late decay to photons coming
from nucleosynthesis. If $\tau \geq 10^4$ s the photonic showers
can change the abundance of light elements \footnote{The
$\gamma-\gamma$ scattering dominates photon interactions
for $t < 10^4$ s and hence photonic showers will not
affect light element abundances.}. For
$10^4 ~ {\rm s} \leq \tau \leq 10^6 ~ {\rm s}$
photodestruction of D will reduce its abundance, while for
$\tau > 10^6$ s photoproduction of D and $^3$He from the destruction
of $^4$He will give the main constraint.

\subsection{$X~(\tilde X) \rightarrow \tilde X~(X) + \gamma + \tilde
\gamma$ via off-shell gravitino}

Henceforth, we shall consider the case
where ${\bf X}$ multiplet has only gravitationally suppressed
couplings to the matter sector\footnote{In the case when $X$ ($\tilde
X$) has common gauge interactions with matter fields the three-body
decays $X_1~(\tilde X) \rightarrow \tilde X~(X_2) + \gamma + {\tilde
\gamma}$ can occur via gauge interactions (in fact through the same
diagrams which results in the production of $X$ ($\tilde X$) in a
thermal bath). In this case the decay rate is quite large $\Gamma \sim
\alpha^2 {\Delta m}^3/{{m_X}^2}$, notice that there is a phase space
suppression, and the decay occurs before nucleosynthesis. Again note that
$X_2$ component is the strictly stable particle.}. Depending on the
mass differences among scalar and fermionic components of $X$
different situations may arise. If $\Delta m \equiv
|m_{X_{1,2}}-m_{\tilde X}| > m_{3/2}$, the decay channels; $X_1
\rightarrow \tilde X + \tilde G$, and $\tilde X \rightarrow X_2 +
\tilde G$ are kinematically allowed. For $\Delta m < m_{3/2}$, and
$\Delta m > m_{\rm
LSP}$, provided the gravitino is not the LSP, the three-body decay
$X_1 \rightarrow \tilde X + \gamma + {\tilde \gamma}$ (and correspondingly
$\tilde X \rightarrow X_2 + {\gamma} + {\tilde \gamma}$) can occur. We
indeed expect that such decays take place since $\Delta m \ll
m_X$ implies that any production mechanism shall produce $X_1$, $X_2$,
and $\tilde X$ (at least $X_1$ and $X_2$) in comparable
abundances. Moreover, we
notice that in both the cases $X_2$ component is the strictly
stable particle and hence a dark matter candidate. Then its abundance,
which is the same as the initial abundance of $X$ ($\tilde X$), must
not saturate the overclosure bound in (9).

\vskip7pt

Let us first consider the case when $\Delta m < m_{3/2}$. This
is the situation in the hidden sector supersymmetry breaking scenario
with the Polonyi field where $\Delta m \approx (1 - \sqrt{3}/2) m_{3/2}$
\cite{keith}. As long as $\Delta m > m_{\rm LSP}$, the
three-body decay $X~(\tilde X) \rightarrow \tilde X~(X) + \gamma +
{\tilde \gamma}$, via an off-shell gravitino, is kinematically
allowed. The decay diagram includes two vertices: at the first vertex
$X$ and ${\tilde X}$ couple to ${\tilde G}$, while gravitino is
coupled to a $\gamma {\tilde \gamma}$ pair at the second vertex.
The decay rate is doubly Planck mass suppressed, and with phase
space suppression leads to

\begin{equation}
\Gamma_1 \sim \frac{1}{32 (2 \pi)^3}\times
\frac{m^{2}_{X} {(\Delta m - m_{\rm LSP})}^{3}}{M^{4}_{\rm p}}\,.
\end{equation}
Here we have assumed that the decay matrix
element is constant over the phase space. An
interesting observation is that the decay rate is model-independent
except for the appearance of $\Delta m$. We make the resonable
assumption that $100~{\rm GeV} \leq \Delta m - m_{\rm LSP} \leq 800$
GeV, for $m_{3/2} \simeq 1$ TeV. Then the lifetime $\tau \sim
\Gamma^{-1}_{1}$ can be of the same
order as the age of the Universe $\simeq
10^{17}$ s for the mass range

\begin{equation}
10^{13} {\rm GeV} < m_X < 10^{15} {\rm GeV}\,.
\end{equation}
For a lighter $m_X$, the lifetime is longer than the age of the
Universe.

\vskip7pt

The decay usually releases entropy, which is worth
estimating in our case. Since $\Delta m - m_{\rm LSP} \ll m_X$, the
released momentum is distributed among all the three decay products
while the energy in the mass difference is practically carried
by $\gamma$ and $\tilde \gamma$. It is therefore reasonable to
consider the energy which is taken away by $\gamma$ to be $({\Delta m
- m_{\rm LSP}})/2$. Then the released entropy is
$s_d \sim ((50-400) n_X)^{3/4}$, while the entropy in
the cosmic microwave background photons is $s \sim n_{\gamma}$,
resulting in
\beq
{s_d \over s} \leq \left(\frac{4 \times 10^{-7}}{T_{\gamma}
m_{X}}\right)^{3/4}\,,
\eeq
where we have used the overclosure bound. Here $T_{\gamma}$ denotes
the temperature of the background photons at the time of decay and has
its smallest value $\sim 5 \times 10^{-13}$ GeV for the decays
occurring at the present moment. This implies that for $m_X \leq 10^6$
GeV, the decay products increase the entropy of the Universe at the
time of decay. However, in this case the decay occurs much later than
today and hence entropy release will not be substantial at present
epoch. Therefore, the three-body decay does not release
any significant entropy within the lifetime of the Universe.

\vskip7pt

Nevertheless, the observed photon flux puts a severe
constraint on the abundance of $X$. For a decay energy of order $(50-400)$
GeV the most stringent limit is \cite{subir1,kr}

\begin{equation}
{n_X \over n_{\gamma}} \leq 10^{-19}\,,
\end{equation}
while for a wide range of decay
lifetimes $10^{17}~{\rm s} \leq \tau \leq 10^{21}~{\rm s}$, the
bound is given by \cite{subir1,kr}
\beq
{n_X \over n_{\gamma}} \leq (10^{-18}-10^{-13})\,.
\eeq
The most conservative mass range which leads to a decay
lifetime in this range is
$10^{11}~{\rm GeV} \leq m_X \leq 10^{17}~{\rm GeV}$. The
overclosure limit now reads $n_X/n_{\gamma} \leq 4 \times 10^{-20}$,
which is more stringent by several orders of magnitude.
Therefore, the three-body decay satisfies bounds on photon flux,
if $\Omega_{X} \leq 0.3$.

\vskip7pt

If $X$ is very massive it will also be
possible that the decay occurs before recombination. For
example, if $m_X \sim 10^{16.5}$ GeV, which can be related to the
string scale in M-theory, and $\Delta m - m_{\rm LSP}
\simeq 1 TeV$, the process
$X \rightarrow \tilde X + \gamma + {\tilde \gamma}$ may take place after
the photon number changing interactions $e~\gamma \rightarrow
e~\gamma~\gamma$
have gone out of equilibrium, but before the recombination era.
For an energy release of order $(50-400)~{\rm GeV}$,
in order to satisfy the present constraint on the chemical
potential, from Eq.~(\ref{const1}), we obtain the bound
$n_{X}/n_{\gamma} \leq 10^{-14}$. This is again much weaker than
the overclosure bound $n_X/n_{\gamma} \leq 4 \times 10^{-25}$. We
therefore conclude that avoiding the overclosure of the Universe
puts a stronger bound on the abundance of $X$ than astrophysical
constraints from the decay process
$X~(\tilde X) \rightarrow {\tilde X}~(X) + \gamma + {\tilde \gamma}$,
via off-shell gravitino.


\subsection{$X~(\tilde X) \rightarrow \tilde X~(X) + \gamma + \tilde \gamma$
via off-shell inflatino}

Now, we turn our attention to another possible decay channel where the
three-body decay of $X(\tilde X)$ can occur via an off-shell inflatino. The
decay diagram will include the $h_X X {\tilde X} {\tilde \phi}$ coupling
at the first vertex while at the second vertex
${\tilde \phi}$ is coupled to a $\gamma {\tilde \gamma}$ pair (as
discussed in the previous section). The significance
of $X$ decay via off-shell inflatino now depends on the dominant
decay channel of inflaton (inflatino). If the inflaton (inflatino)
dominantly decays to a three body final state, we will have a
four-body $X$ decay. This is even more phase space suppressed (and perhaps
forbidden) than the three-body decay via off-shell gravitino. On the
other hand, if the inflaton (inflatino) mainly decays to two body final state,
in particular to a $\gamma {\tilde \gamma}$ pair, we may have a
three-body decay of $X$ via the inflatino. Then the rate for the
decay $X~(\tilde X) \rightarrow {\tilde X}~(X) + \gamma + {\tilde
\gamma}$ is found to be

\begin{equation}
\label{decay}
\Gamma_2 \sim 10^{-2} h^{2}_{X} {T^{2}_{\rm R}
M_{\rm Planck} \over m^{3}_{\phi}} {{(\Delta m - m_{\rm LSP})}^3 \over
M^{2}_{\rm Planck}}\,.
\end{equation}
As pointed out earlier, the coupling $h_X$ must satisfy the inequality
in (8) in order for $X$ not to overclose the Universe, if $m_X
<m_{\phi}/2$. This yields

\begin{equation}
\Gamma_2 \sim {4 \times 10^{-9} \over m_X}
{\left ({T_{\rm R} \over m_{\phi}} \right )}^3 {(\Delta m - m_{\rm LSP})^3
\over M^2_{\rm Planck}}\,.
\end{equation}

Let us first consider the case with a decay energy of order $400$
GeV. Then for $m_X = 10^5~(10^{10})~{\rm GeV}$ the decay lifetime
is $\tau_2 \sim 10^{17}~(10^{22})~{\rm s}$,
provided $T_{\rm R} \simeq m_\phi$. The gamma-ray background
and the overclosure bounds then translate to
$n_X/n_\gamma \leq 10^{-17}~(10^{-12})$ \cite{subir1,kr}, and
$n_X/n_\gamma \leq 10^{-14}~(10^{-19})$, respectively. For masses
$10^5~{\rm GeV} \leq m_{X} \leq 10^7~{\rm GeV}$ the gamma-ray bound
is indeed stronger. Therefore, providing a clear signal that
within this mass range $X$ can not act as a stable dark matter candidate.
For $T_{\rm R}/m_\phi \simeq 10^{-1}$ astrophysical constraints
turn out to be more stringent than the overclosure bound within the
mass range
$10^3~{\rm GeV} \leq m_X \leq 10^5~{\rm GeV}$.
While for a smaller $T_{\rm R}/m_\phi$, the main
constraint appears from the overclosure bound. For example, for
$T_R/m_\phi \leq 10^{-2}$ the overclosure bound ensures that for $m_X
\geq 1$ TeV the three-body decay will not be atsrophysically dangerous.

For a decay energy of order $50$
GeV a mass $m_X = 10^3~(10^7)~{\rm GeV}$ results in the decay lifetime
$\tau_2 \sim 10^{18}~(10^{22})~{\rm s}$,
provided $T_{\rm R} \simeq m_\phi$. The gamma-ray background
and the overclosure bounds yield
$n_X/n_\gamma \leq 10^{-18}~(10^{-13})$ \cite{subir1,kr}, and
$n_X/n_\gamma \leq 10^{-12}~(10^{-16})$, respectively, where for
$10^3~{\rm GeV} \leq m_{X} \leq 10^5~{\rm GeV}$ the gamma ray bound
will be stronger. For $T_{\rm R}/m_\phi \simeq 10^{-1}$ astrophysical
constraints are more stringent within a narrow mass range $10^3~{\rm
GeV} \leq m_X \leq 10^4~{\rm GeV}$, while for a smaller $T_{\rm
R}/m_\phi$ the main constraint appears from the overclosure bound.

\vskip7pt

If we desire to have a supermassive $X$ with a mass  $m_X > m_\phi/2$, the
production of $X$ through direct decay of the inflaton cannot be
possible. Therefore, even if such massive $X$ might have been created, their
decay would follow Eq.~(\ref{decay}) and could easily occur before
recombination. However, mechanisms which could create $X$ in this case prefer
a mass $m_X > 10^{10}$ GeV \cite{ckr1,ckr2,cckr}. The
overclosure bound then yields $n_{X}/n_{\gamma} \leq
10^{-19}$, which for an energy release $50-400$ GeV in the three-body
decay is always stronger than the astrophysical constraints.

Our overall conclusion is that for
$m_{3/2} \simeq 1$ TeV the decay process $X~(\tilde X) \rightarrow {\tilde
X}~(X) + \gamma + {\tilde \gamma}$ can be astrophysically dangerous when
mediated by inflatino.  In particular, the astrophysical bounds are
stronger than the overclosure bound for the mass range $10^3~{\rm
GeV} \leq m_X \leq 10^7~{\rm GeV}$, when $10^{-1} \leq T_R/m_\phi \leq
1$.

\subsection{$X~({\tilde X}) \rightarrow {\tilde X}~(X)+{\tilde G}$}

Let us finish by discussing the two-body decay case which occurs when
$\Delta m > m_{3/2}$. This is also quite plausible, since, as
pointed out in Ref.~\cite{keith}, for a generic
supersymmetry breaking scenario one may not have $\Delta m < m_{3/2}$.
The decay rate for the process
$X~({\tilde X}) \rightarrow {\tilde X}~(X)+{\tilde G}$ is however
Planck mass suppressed, and phase space suppression gives
\beq
\Gamma_3 \sim {m^{2}_{X}m_{3/2} \over 8 \pi M^{2}_{\rm
Planck}}.
\eeq
As we have mentioned earlier, in such a case the most important bounds
come from the success of big bang nucleosynthesis.
Here the main danger arises from the produced gravitinos with a mass
$m_{3/2} = 100$ GeV, which decay into energetic photons thus
constraining their abundance to $n_X/n_{\gamma} \leq 10^{-14}$
\cite{subir} (here we obviously assume that the gravitino is not the
LSP). This requires that $m_X \geq 10^6$ GeV, if $X$ is produced with an
abundance $\Omega_{X} \sim 0.3$. On the other hand, for a
gravitino mass $m_{3/2} = 1$ TeV, the nucleosynthesis bound results in
$n_X/n_{\gamma} \leq 10^{-12}$ \cite{subir}, which relaxes the mass
range of $X$ and masses beyond $10$ TeV could easily be accommodated.

\vskip7pt

Some comments are in order. The decay of a particle into its
superpartner and a gravitino can in general occur for unstable
species as well. However, in such cases the suppression of this decay mode
relative to other decay channels ensures that nucleosynthsis will not
be disrupted (e.g. the case for inflaton decay has been considered in
Ref. \cite{keith}. On the other hand, for a particle which is stable
in the limit of unbroken supersymmetry, this decay mode (and the three-body
decays discussed earlier) are the only possible channels. We also
limited our discussion to gravity-mediated models of supersymmtery
breaking. The reason is that in gauge-mediated models the gravitino
mass is substantialy samller than the weak scale implying that the
gravitino is
the LSP. Then a possible two-body decay does not pose a
threat to nucleosynthsis, while three-body decays are kinematically
forbidden. Finally, we only considered neutral stable particles in
this paper. For a charged particle, the dominant decay mode will be
$X~(\tilde X) \rightarrow {\tilde X}~(X) + {\tilde \gamma}$ which occurs much
before nucleosynthsis. Moreover, the abundance of such particles is
severely constrained by their searches in the sea water which is much
more stringent than the overclosure bound \cite{ky}.

\section{Conclusion}

We have discussed the possible astrophysical signatures of
a neutral stable particle within the context of supergravity. We
considered a multiplet $\bf X$ which has a mass $m_X$ and is stable in
the limit of unbroken
supersymmtery. Supersymmtery breaking in the hidden sector
generally results in the pattern $m_{X_1} > m_{\tilde X} > m_{X_2}$, so
long as $m_X \gg m_{3/2}$, where $X_1$, $X_2$, and $\tilde X$ are the two
scalar components and the fermionic component of the multiplet
respectively. We have noticed that a gravitationally suppressed
inflaton decay could lead to production of $X$ ($\tilde X$) in
interesting abundances in a wide mass range $10^3~{\rm GeV} \leq
10^{11}$ GeV. Details depend on the inflaton VEV at the
minimum and the reheat temperature of the Universe, which we always
assume to be smaller than $10^{10}$ GeV, in order to avoid thermal
gravitino overproduction.

\vskip7pt

We also considered the decay of $X$ ($\tilde
X$) into its superpartner and a photon-photino pair, or a
gravitino, and discussed the various astrophysical and
overclosure bounds which restrict the decay channels. If $X$ ($\tilde
X$)is
produced with an abundance large enough to account for the
dark matter in the Universe, such decays merely change the identity of
the dark matter candidate. Depending on the supersymmetry breaking
scenario, the mass difference between $X$ and $\tilde X$ can be smaller
or larger than the mass of the gravitino. In the
former case, the two-body decay channels are kinematically forbidden.
However, the three-body decay via off-shell gravitino and/or off-shell
inflatino may still occur. The three-body decays release an energy
which is just the mass difference between $X$, $\tilde X$ and LSP.
In the case where $X$ ($\tilde X$) decays into $\tilde X$ ($X$) and a
photon-photino pair, via an off-shell gravitino, for the mass range
$10^{11}{\rm GeV} \leq m_X \leq 10^{17}$ GeV
the main constraint comes from the overclosure bound yielding
$n_{X}/n_{\gamma} \leq 10^{-20}$. For $X$ ($\tilde X$) decaying via
off-shell
inflatino the situation depends on a number of model parameters such
as the reheat temperature, and mass of the inflaton.

We also pointed out
that for larger ratios of $T_{\rm R}/m_\phi$ the gamma ray background
could in principle constrain the lower half of the mass range
$10^3~{\rm GeV} \leq m_X \leq 10^7~{\rm GeV}$, while the upper half is
constrained by the overclosure limit. We also considered the decay of
$X$ ($\tilde X$)
to $\tilde X$ ($X$) and a gravitino when the mass difference between them is
larger than the gravitino mass. In this case  the main constraint comes from
avoiding the photodissociation of light elements. Indeed, the
abundance of $X$ must be smaller than $10^{-14}$ for $m_{3/2} = 100$
GeV implying that $m_X \geq 10^6$ GeV, if $X$ is the dark
matter in the Universe.

\section{acknowledgements}
The authors are thankful to A. {\"O}zpineci, A. Perez-Lorenzana, and
S. Sarkar for valuable discussions. R.A. is supported by
``Sonder-forchschungsberich 375 f$\ddot{\rm u}$r Astro-Teilchenphysik''
der Deutschen Forschungsgemeinschaft, K.E. partly by the
Academy of Finland  under  the contract 101-35224, and A.M. acknowledges the
support of {\it The Early Universe network} HPRN-CT-2000-00152, and
the hospitality of the Helsinki Institute of Physics where part
of the work has been carried out.



\begin{references}

\bibitem{nath}P. Nath, R Arnowitt and A. Chamseddine,
{\it Applied N=1 Supergravity} Singapore, World Scientific, (1984);
R. Barbieri, S. Ferrara and C.A. Savoy, Phys. Lett. {\bf B119}, 343 (1982);
A. Chamseddine, R. Arnowitt and P. Nath, Phys. Rev. Lett. {\bf 49}, 970
(1982); H.P. Nilles, M. Srednicki and D. Wyler, Phys. Lett. {\bf B120},
346 (19983); L. Alvarez-Gaumez, J. Polchinski and M. Wise,
Nucl. Phys. {\bf B221}, 495 (1983).

\bibitem{deser}
S. Deser and B. Zumino, Phys. Rev. Lett. {\bf 38}, 1433 (1977).

\bibitem{ellis}
J. Ellis, J.E. Kim, and D.V. Nanopoulos, Phys. Lett. {\bf B145}, 181 (1984);
J. Ellis, D.V. Nanopoulos, K.A. Olive and S-J. Rey, Astropart. Phys.
{\bf 4}, 371 (1996),\\
for a recent calculation see: M. Boltz,
A. Brandenburg and W. Buchmuller, Nucl. Phys. {\bf B606}, 518 (2001).

\bibitem{maroto}
A. L. Maroto and A. Mazumdar, Phys. Rev. Lett {\bf 84}, 1655 (2000).

\bibitem{kallosh}
R. Kallosh, L. Kofman, A. Linde and A. Von Proeyen,
Phys. Rev D {\bf 61}, 103503 (2000); G. F. Giudice, I. Tkachev and
A. Riotto, JHEP {\bf 9908}, 009 (1999).

\bibitem{rouzbeh}
R. Allahverdi, M. Bastero-Gil, and A. Mazumdar, Phys. Rev. D. {\bf 64},
023516 (2001).

\bibitem{nps}H.P. Nilles, M. Peloso and L. Sorbo,
Phys. Rev. Lett. {\bf 87}, 051302 (2001); JHEP 0104, 004 (2001).

\bibitem{subir}M. Kawasaki, K. Kohri and T. Moroi, Phys. Rev. {\bf
D63}, 103502 (2001);\\
for a review see: S. Sarkar, Rept. Prog. Phys. {\bf 59}, 1493 (1996).

\bibitem{beatriz}
G.D. Coughlan, W. Fischler, E.W. Kolb, S. Raby and
G.G. Ross, Phys. Lett. {\bf B131}, 53 (1983); J. Ellis, D.V. Nanopoulos,
M. Quiros, Phys. Lett. {\bf B174}, 176 (1986);
B. de Carlos, J.A. Casas, F. Quevedo, E. Roulet, Phys. Lett. {\bf
B318}, 447 (1993).

\bibitem{kt}E.W. Kolb, and M.S. Turner, {\it The Early Universe},
Addison Wisley, 1990.

\bibitem{gk} K. Griest and M. Kamionkowski, Phys. Rev. Lett. {\bf 64},
615 (1990).

\bibitem{ckr1}D.J.H. Chung, E.W. Kolb and A. Riotto, Phys. Rev. {\bf
D59}, 023501 (1999).

\bibitem{ckr2}D.J.H. Chung, E.W. Kolb and A. Riotto, Phys. Rev. Lett{\bf
81}, 4048 (1998).

\bibitem{ckr3}D.J.H. Chung, E.W. Kolb and A. Riotto, Phys. Rev. {\bf
D60}, 063504 (1999).

\bibitem{cckr}D.J.H. Chung, P. Crotty, E.W. Kolb and A. Riotto, Phys. Rev. {\bf
D64}, 043503 (2001).

\bibitem{subir1}J. Ellis, G.B. Gelmini, J.L. Lopez, D.V. Nanopoulos
and S. Sarkar, Nucl. Phys. {\bf B373}, 399 (1992).

\bibitem{kr}G. Kribs and I. Rothstein, Phys. Rev. {\bf D55}, 4435
(1997); erratum Phys. Rev. {\bf D56}, 1822 (1997).

\bibitem{dolgov}A.D. Dolgov and Ya.B. Zeldovich, Rev. Mod. Phys. {\bf 53},
3 (1981); P.J.E. Peebles, {\it Physical Cosmology}, Princeton University
Press, 1971.

\bibitem{mather}J.C. Mather, $et~ al.$, Ap. J. {\bf 420}, 439 (1994).

\bibitem{keith}H.P. Nilles, K.A. Olive and M. Peloso, hep-ph/0107212.

\bibitem{ky}A. Kudo and M. Yamaguchi, Phys. Lett. {\bf B516}, 151 (2001).

\end{references}
\end{document}